\documentclass[12pt]{article}

\usepackage{newtxtext,newtxmath}
\usepackage[T1]{fontenc}
\usepackage{caption}
\usepackage{graphicx}
\usepackage{float}

\usepackage[letterpaper,margin=1in]{geometry}

\renewenvironment{abstract}
	{\quotation}
	{\endquotation}

\date{}

\usepackage{listings}
\makeatletter
\renewcommand{\fnum@figure}{\textbf{Figure \thefigure}}
\renewcommand{\fnum@table}{\textbf{Table \thetable}}
\makeatother

\usepackage{scicite}
\usepackage{placeins}
\usepackage{url}

\def\scititle{
Solid-State NMR Dipolar Recoupling in Presence of Large Chemical Shielding Anisotropies  by Quaternion-Based Effective Hamiltonian  Optimal Control
}
\title{\bfseries \boldmath \scititle}

\author{
	Enik\H{o} Baligács, Nino Wili, José P. Carvalho,
	Anders Bodholt Nielsen,$^{\ast}$
	 \\ and
    Niels Chr. Nielsen$^{\ast}$
 \and
	\small Interdisciplinary Nanoscience Center (iNANO) and Department of Chemistry, Aarhus University, 
    \\ \small Gustav Wieds Vej 14, DK-8000 Aarhus C, Denmark. 
    \\
	\small$^\ast$Corresponding authors. Email: ncn@chem.au.dk and abn@chem.au.dk \and
}

\begin{document} 

\maketitle

\begin{abstract} \bfseries \boldmath
\footnotesize
 Dipolar recoupling is a key element in magic-angle-spinning (MAS) solid-state NMR spectroscopy with reintroduced dipole-dipole coupling interactions providing information about internuclear distances and enabling transfer of polarization between spins in resolution-enhancing multiple-dimensional experiments. Such methods may be challenged in many important applications by the presence of large anisotropic nuclear spin interactions such as chemical shielding anisotropy. In this paper, we address this challenge by presenting quaternion-based optimal control.  This is founded in single-spin operations enabling optimization of effective Hamiltonians with reduced influences from anisotropic shielding. Along with the principles underlying such optimizations, we present numerical and experimental demonstration of $^{19}$F to $^{13}$C polarization transfer in presence of $^{19}$F chemical shielding anisotropy.
\end{abstract}

\newpage
\noindent

\section{Introduction}

Dipolar recoupling \cite{Nielsen2012,Ji:2021aa,ladizhansky2024DipolarRecoupling} constitutes a key ingredient in modern magic-angle-spinning (MAS) solid-state NMR experiments, enabling the controlled reintroduction of dipole-dipole coupling interactions that are otherwise averaged by sample rotation. Reintroduced dipole-dipole couplings may provide direct information about structurally important internuclear distances, molecular dynamics, and serve as means to transfer polarization  between nuclear spins aimed at enhancement of resolution in multiple-dimensional experiments. The application of dipolar recoupling in solid-state NMR spans very wide,\cite{Nielsen2012,Ji:2021aa,ladizhansky2024DipolarRecoupling} with recent representative examples including structural studies of biological systems\cite{lasorsa2025solid, Reif:2021aa,Ahlawat:2022aa} and investigations of functional materials.\cite{gao2025solid, zhang2024elucidations, haro2024solid,ssNMRMaterialsRev2020,Reif:2021aa,Ahlawat:2022aa} 

A large variety of dipolar recoupling schemes has been developed to meet different demands relating to instrumentation, experimental parameters, mode and aim of recoupling, and details of the involved spin systems.\cite{Nielsen2012,Ji:2021aa,ladizhansky2024DipolarRecoupling} Pioneered by relatively simple recoupling methods like rotational resonance,\cite{R2} rotational echo double resonance (REDOR),\cite{gullion1989REDOR} double cross polarization (DCP), \cite{DCP}  homonuclear rotary resonance (HORROR)\cite{nielsen1994HORROR}, and radio-frequency driven recoupling (RFDR),\cite{bennett1992RFDR} recoupling gradually involved increasingly advanced and powerful pulse sequences such as a large variety of symmetry-based recoupling experiments.\cite{C7,POSTC7,carravetta2000symmetry,Levitt_symm_rec} Such experiments as well as optimal control derived pulse sequences\cite{OCsolid,Nielsen2010_OC_encycl} may offer improved offset compensation, interaction specificity,  better match between radio-frequency (RF) irradiation and sample spinning, and robustness to instrumental imperfections such as RF inhomogeneity and phase transients. 

Despite these advances, the use of dipolar recoupling for heteronuclear polarization transfer remains challenging in many systems of high practical relevance. This applies in particular for spin systems influenced by large anisotropic nuclear spin interactions such as chemical shielding anisotropy (CSA) and quadrupolar coupling interactions. Such sample-rotation modulated interactions, often being much larger than the dipole-dipole couplings intended for recoupling, may severely distort establishment of required resonances between sample spinning and radio-frequency irradiation and significantly reduce efficiency of coherence or polarization transfer.\cite{odgaard200113c,de2012dipolar}
While established approaches such as  ramped or adiabatic variants of cross polarization, and symmetry-based sequences can partially mitigate these effects, they often rely on relatively narrow matching conditions or exhibit pronounced sensitivity to the combined influence of offset and CSA.  In context of homonuclear dipolar recoupling, asynchronous implementations of C7\cite{C7,POSTC7} type pulse squences has been proposed.\cite{Asynchron_C7} In the context of heteronuclear dipolar recoupling, frequency-swept and adiabatic cross-polarization approaches such as BRAIN-CP-MAS\cite{wi2015BRAINCPMAS} have demonstrated remarkably broadband polarization transfer under MAS, particularly for systems with a large span of  isotropic chemical shift offset and large anisotropic shielding interactions. Indirect-detection methods relying on residual $^1$H dipolar order have likewise been shown to provide efficient broadband polarization transfer in systems with large chemical-shielding anisotropies under MAS \cite{wolf2024sensitivity}.  Recently, optimal control pulse sequences such as   OPTIANS \cite{OC_ssnmr_complex_jain} have been introduced to cope with large anisotropic interactions in heteronuclear cross-polarization experiments. Each of these approaches  solves specific challenges, but a more general approach to derive optimal dipolar recoupling experiments in presence of dominant anisotropic shielding interactions remains a need to expand dipolar recoupling to a large range applications in materials and biological sciences.

In this work, we address the challenge  of dipolar recoupling in presence of large CSA's from a methodology point of view by combining single-spin-vector effective Hamiltonian theory (SSV-EHT)\cite{shankar,SSV-EHT,SVEHT_EEHT} with  optimal control procedures explicitly taking into account dominant isotropic and anisotropic chemical shift interactions to ensure convergence of effective Hamiltonians to first order. We do this from the philosophy that new design strategies may be beneficial in the quest to solve challenges not coped sufficiently well with using existing methods. Within this framework, the optimization challenge is further simplified significantly through formulation at the level of convergent first-order effective single-spin Hamiltonians active over short modulation periods (i.e, short periodic pulse sequences). The reduction of the problem to single-spin operation enables  the use of an efficient quaternion-based\cite{Hamilton1844,Quarternions} optimal control optimization of the controlling effective field.
Using this approach, we derive a pulse sequence element that generate stable and well-defined effective fields across  a broad range of isotropic and anisotropic shielding interactions. It is further demonstrated how individual pulse sequence elements can generate a large variety of recoupling experiments through combination with different pulse shapes on the second channel. This provides a systematic route to robust heteronuclear dipolar recoupling in regimes where conventional approaches are challenged, and establishes a flexible design strategy for different characteristics on both spins individually to  achieve desired recoupling conditions.

\FloatBarrier
\section{Theory}

In this section, we outline the theoretical framework underlying the optimization of pulse elements for compensation of large isotropic and anisotropic shielding effects during heteronuclear dipolar recoupling. The approach, which combines single-spin-vector effective Hamiltonian theory (SSV-EHT),\cite{SVEHT_EEHT} quaternion algebra,\cite{Quarternions} and gradient-based optimal control,\cite{grape,OCsolid} may conveniently be described in a series of steps.

We consider the time-dependent Hamiltonian of a heteronuclear I-S spin system under MAS conditions, taking into account RF irradiation, isotropic and anisotropic shifts on both spins, and  heteronuclear dipole-dipole coupling between the  I and S spins
\begin{equation}
\mathcal{H}(t) = \mathcal{H}_\text{I}(t)+\mathcal{H}_\text{S}(t)+\mathcal{H}_\text{IS}(t) \quad ,
\end{equation}
with the linear (single-spin, I and S) and bilinear (two-spin, IS) Hamiltonians defined as
\begin{eqnarray}
        \mathcal{H}_\text{K}(t) &=& \sum_{m=-2}^{2} \omega_{\text{K}}^{(m)} e^{i  m\omega_rt}K_z \nonumber \\&+&\omega_\text{RF}^\text{K}(t) [\cos\phi^{\text{K}}_{\text{RF}}(t) K_x + \sin\phi^{\text{K}}_{\text{RF}}(t)K_y] \\
      \mathcal{H}_\text{IS}(t)  &=&\sum_{m=-2}^2 \omega_\text{IS}^{(m)}e^{im\omega_rt}2I_zS_z ,
\label{eqn:Ham}
\end{eqnarray}
where
 $\omega^\text{K}_{\text{RF}}(t)$ and $\phi^{\text{K}}_{\text{RF}}(t)$ represents the time-dependent RF amplitude and phase on spin K (K = I or S), respectively, and $\omega_r$  the sample spinning frequency. $\omega_{\text{K}}^{(m)}$ and $\omega_{\text{IS}}^{(m)}$ are the Fourier components of the K-spin shielding and IS  dipole-dipole coupling interactions
\begin{eqnarray}  
      \omega_\text{K}^{(m)} &=& \omega_\text{K}^\mathrm{iso} \delta_{m,0} + 
	  \omega_\text{K}^\mathrm{aniso} \{
	  D^{(2)}_{0,-m}(\Omega_{PR}^\text{K}) \nonumber \\
	  &-& \frac{\eta_\text{K}}{\sqrt{6}}[
	  D^{(2)}_{-2,-m}(\Omega_{PR}^\text{K})+D^{(2)}_{2,-m}(\Omega_{PR}^\text{K})
	  ]
	  \} d^{(2)}_{-m,0}(\beta_{RL}) , \nonumber \\
	  \omega_\mathrm{IS}^{(m)} &=& b_\mathrm{IS} D^{(2)}_{0,-m}(\Omega_{PR}^\mathrm{IS})
	  d^{(2)}_{-m,0}(\beta_{RL}) ,
\label{eq:rotCS2}  
\end{eqnarray}
where $\delta_{m,0}$ is the Kronecker delta (equal to 1 for $m=0$, 
otherwise 0) and $\omega_\text{K}^\mathrm{iso}$ and $\omega_\text{K}^\mathrm{aniso}$ represent the isotropic and anisotropic chemical shift, and $b_\mathrm{IS}$ the dipole-dipole coupling constant. $D^{(2)}_{m',m}$ denotes the $m'$, $m$ component of a second-rank Wigner matrix with $\Omega_{PR}^\text{M}=\{\alpha_{PR}^\text{M},\beta_{PR}^\text{M},\gamma_{PR}^\text{M}\}$ being Euler angles transforming the principal axis frame ($P$) spatial tensor of the interaction M (M=I, S, IS) into the rotor frame ($R$) and $\beta_{PL}$ the angle between the rotor axis and the laboratory frame ($L$) longitudinal axis.\cite{SIMPSON,Nielsen2012} All frequencies are in angular units.
This Hamiltonian is completely general and will in the context of quaternion optimization be treated with separate rotations on the I and S spins. For convenience, we initially consider a spin system with anisotropic shielding on the I spin only.  The isotropic shift will be treated as  offset relative to the RF carrier frequencies, i.e., $\Delta \omega_{\text{K}}^\mathrm{iso}$=$\omega_{\text{K}}^{\text{iso}}-\omega^\text{K}_\text{RF,carrier}$. 

Following the SSV-EHT formalism,\cite{SVEHT_EEHT} we apply a sequence of frame transformations involving single-spin transformations on each of the two spins to set focus on and optimize dipolar recoupling. The latter is mediated by bilinear terms formed by products of transformed single-spin operators. This favorably involve consideration of isotropic and anisotropic shielding as well as RF pulse irradiation on both spins, or may involve part of this, e.g., only  anisotropic shielding and RF irradiation on the I spin with subsequent matching with an independent RF pulse sequence on the other S spin. The latter will be a focal point in this paper. In a first step, we transform in the most general Hamiltonian into the interaction frame defined by the isotropic and anisotropic shielding and the RF pulse sequences (synchronized) for the two spins. This transformation demodulates these contributions from the Hamiltonian through the corresponding Coriolis terms. Notably, in contrast to previous SSV-EHT formulations, the  anisotropic shielding interaction is explicitly included in this step, adopting an idea presented in Refs. \cite{CARAVATTI198388,10.1063/1.3521491}, such that its time-modulated effect is fully accounted for in the transformed frame. We note that including isotropic as well as anisotropic shifts, in addition to RF irradiation, into the interaction frame transformation markedly improves convergence of effective Hamiltonian averaging - with the first-order term efficiently catching dominant parts of the spin dynamics.

In a second step, periodicity with respect to the modulation frequency $\omega_m$ of the RF irradiation ($\tau_m=2\pi/\omega_m$ is the length of the pulse sequence element) is encoded for the remaining dipole-dipole coupling interaction. In combination these steps introduce constant effective fields, $\omega_\text{eff}^{\text{K}}$, on the two spins. Finally, the description is transformed into a tilted frame with these two effective fields aligned along the $z$, i.e., $\tilde{I}_z$ and $\tilde{S}_z$ for the two spins individually. The leads to formulation of the residual dipolar coupling as a  combined Fourier expansion involving  $\omega_r$ and  $\omega_m$ as
\begin{multline}
\tilde{\mathcal{H}}(t) = -\omega_{\text{eff}}^{\text{I}}\tilde{I}_z - \omega_{\text{eff}}^{\text{S}}\tilde{S}_z 
+ \sum_{m=-2}^{2} \sum_{p,q=x,y,z} \sum_{k_\text{I},k_\text{S}=-\infty}^{\infty}
\omega_{\text{IS}}^{(m)} e^{im\omega_r t} 
a_{k_\text{I}}^p e^{ik_\text{I}\omega_m t}
a_{k_\text{S}}^q e^{ik_\text{S}\omega_m t}
\, 2 \tilde{I}_p \tilde{S}_q ,
\label{eqn:Ham_SSV}
\end{multline}
where $a_{k_\text{I}}^p$ and $a_{k_\text{S}}^q$ are Fourier coefficients describing the periodic modulations induced by the RF irradiation with $k_\text{I}$, $k_\text{S}$ being integer Fourier indices.

Polarization transfer is achieved by the matching of the effective linear fields on the RF channels on the two spins implying
\begin{equation}
\omega_{\text{eff}}^{\text{I}} = \pm \omega_{\text{eff}}^{\text{S}}
\label{eqn:HeffCondition}
\end{equation}
with $-$ and $+$ referring to zero- (ZQ) or double-quantum (DQ) match, respectively. This condition must be fulfilled over  the desired range of isotropic and anisotropic shieldings on the I spin.\cite{baligacs2026Accordion, DNP_SSNMR_Carvalho:2025aa} This requires that the linear terms do not truncate the relevant bilinear operators, being a key target in the  optimization described in the next section.

\subsection{Quaternion Optimal Control}

Within the SSV-EHT formulation, the optimization problem reduces to the design of a single-spin effective field trajectory over one modulation period $\tau_m$.
Since the linear effective-field properties of the two spin channels can be optimized largely independently (Hartmann-Hahn match requires Eq.~(\ref{eqn:HeffCondition}) to be fulfilled, while the pulse-sequence optimization may be carried out at the single-spin level). 
This is valid as long as the linear terms in Eq.~(2) are larger than the bilinear terms in Eq.~(3). 
The optimization therefore reduces to designing a pulse sequence that generates the desired effective field, $\omega_\mathrm{eff}^{\mathrm{I}}$, while explicitly accounting for isotropic offset and chemical shielding anisotropy. To efficiently propagate the spin dynamics on level of effective Hamiltonians, we employ the well-established quaternion representation of effective fields and rotations.\cite{Hamilton1844,Quarternions}

For a pulse sequence consisting of $n$ piecewise-constant intervals, the Hamiltonian at the $j$th step is considered time independent, with duration $\Delta t_j=t_j-t_{j-1}$. The RF field is characterized by its amplitude $\omega_{\text{RF},j}$=$\omega_{\text{RF}}(t_j)$, phase $\phi_{\text{RF},j}$=$\phi_{\text{RF}}(t_j)$, and offset $\Delta\omega_{\Omega, j}$=$\Delta \omega^\text{iso}+\omega^\text{aniso}(\Omega_{PR},\beta_{RL}, \omega_{r}t_j)$ (i.e., $\Omega$ includes the powder $\Omega_{PR}$ and $\beta_{RL}$ functionality in Eq. (2)).  The corresponding  effective field is 
\begin{equation}
|\omega_{j,\Omega}| = \sqrt{\omega_{\mathrm{RF},j}^2 + \Delta\omega_{\Omega, j}^2} \quad .
\end{equation}
The rotation during the $j$th interval is represented by a four-component quaternion vector, analogous to the real-valued representation used by Blümich and Spiess,\cite{Quarternions}
\begin{align}
\vec{q}_{j,\Omega} =\left[
\begin{aligned}
&~~~\cos\!\left(\frac{|\omega_{j,\Omega}|\Delta t_j}{2}\right) \\
&-\sin\!\left(\frac{|\omega_{j,\Omega}|\Delta t_j}{2}\right)
\frac{\omega_{\text{RF},j}}{|\omega_{j,\Omega}|}\cos\phi_{\text{RF},j} \\
&-\sin\!\left(\frac{|\omega_{j,\Omega}|\Delta t_j}{2}\right)
\frac{\omega_{\text{RF},j}}{|\omega_{j,\Omega}|}\sin\phi_{\text{RF},j} \\
&-\sin\!\left(\frac{|\omega_{j,\Omega}|\Delta t_j}{2}\right)
\frac{\Delta\omega_\Omega}{|\omega_{j,\Omega}|}
\end{aligned}
\right].
\end{align}
The present work employs a scalar-first quaternion convention instead of the mathematically equivalent scalar-last convention used in Ref.~\cite{Quarternions}; the equivalence of the two conventions is outlined in the Supplementary  Information.

The quaternion describing the rotation of the entire pulse sequence, $\vec{q}_{\mathrm{total},\Omega}$, is obtained by successive evolution of the quaternions representing the individual pulse elements,
\begin{equation}
\vec{q}_{\mathrm{total},\Omega}=\vec{q}_{n,\Omega}\otimes\vec{q}_{n-1,\Omega}\otimes\cdots\otimes\vec{q}_{1,\Omega} \quad ,
\end{equation}
where $\vec{q}_{j,\Omega}$ denotes the quaternion corresponding to the $j$th pulse element and $\otimes$ denotes quaternion multiplication for which the equivalent matrix--vector formulation is derived in the Supplementary  Information.\cite{Quarternions,Voight2021QuaternionAlgebra}

The optimization target is defined by the fidelity function, which measures the overlap between the propagated quaternion, $\vec q_{\mathrm{total},\Omega}$, and the desired target rotation quaternion, $\vec{q}_\text{target}$, as
\begin{equation}
\Phi = \frac{1}{n_\Omega n_{\Delta\omega^\mathrm{iso}}}\sum_{\Omega} \sum_{\Delta\omega^\text{iso}}
\vec{q}_\text{target}^T \vec{q}_{\mathrm{total},\Omega} \quad , 
\label{eqn:fidelityFunction}
\end{equation}
where $n_\Omega$ and $n_{\Delta\omega^\mathrm{iso}}$ denote the total numbers of crystallite orientations and isotropic chemical shift points, respectively.

Gradients with respect to the pulse parameters at each time-step are evaluated analytically. Since the total quaternion is constructed as a product of the individual time-step quaternions, differentiation follows directly from the product rule. 
For the optimization parameter $p_j \in \{\omega_{\mathrm{RF},j},\,\phi_{\mathrm{RF},j}\}$,  associated with time step $j$, the analytical gradient may be expressed as
\begin{equation}
\frac{\partial \Phi}{\partial p_j} = \vec q_\mathrm{target}^T\cdot
\left(\vec{q}_{n,\Omega}\otimes...\otimes \vec{q}_{j+1,\Omega} \otimes \frac{\partial \vec q_{j,\Omega}}{\partial p_j} \otimes
\vec{q}_{j-1,\Omega}\otimes...\otimes \vec{q}_{1,\Omega}\right).
\label{eqn:gradient}
\end{equation}
More details on the gradient can be found in the Supplementary Information.
In similarity to the GRAPE optimal control algorithm,\cite{grape} intermediate propagation results are stored, enabling efficient evaluation of all gradients without repeated propagation throughout the complete pulse sequence.


\FloatBarrier

\section{Experimental and Computational Details}
\subsection{Optimization and Single-Spin Analysis}
Optimization of the pulse sequences, henceforth referred to as Quaternion Optimize (QOpt) pulse sequences, was performed based on the formalism outlined above and using an in-house quaternion-based optimal control program implemented in MATLAB (see source code availability). The optimization employed MATLAB's \texttt{fmincon} solver\cite{MATLAB_fmincon} with analytically evaluated objective-function gradients (cf., Eq.~(\ref{eqn:gradient})), enabling efficient gradient-based optimization. The optimization variables were the piecewise-constant RF amplitudes and phases of the pulse sequence element applied on the relevant (I-spin) RF channel. The RF amplitudes were constrained to $0\leq\omega_\text{RF}/(2\pi)\leq100 $~kHz, while the RF phases were optimized without restriction. The objective function was the offset-averaged transfer fidelity (cf., Eq.~(\ref{eqn:fidelityFunction})), simultaneously considering I-spin isotropic chemical shift offsets and chemical shielding anisotropy to obtain pulse sequences that are robust toward both interactions during I$\rightarrow$S cross polarization.

In all cases, the target was a $120^\circ$ rotation about the $x$ axis optimized using a pulse sequence element with a modulation time of $\tau_m = 80~\mu$s  discretized into $n$ = 40 elements.  The optimization was carried out over an isotropic offset range of 70 kHz (variation of the RF carrier frequency relative to the isotropic chemical shift) using 7 linearly spaced sampling points, with a maximum of 10\,000 iterations. 
The rotor period was set to $\tau_r$ = 40~\textmu s, corresponding to a MAS frequency of $\omega_r/(2\pi)$ = 25 kHz. 
Powder averaging was performed using 5 $\gamma_{CR}$ angles and 10 $(\alpha_{CR}, \beta_{CR})$ REPULSION angles.\cite{REPULSION} 
The chemical shift anisotropy were set to $\omega^\text{aniso}/(2\pi)$ = 80 kHz and the asymmetry parameter to $\eta = 0.5$.

\subsection{Two-Spin Simulations}

All two-spin simulations (Fig.~\ref{fig:fig2}) were performed using SIMPSON.\cite{SIMPSON,SIMPSON-PP} 
The simulations assumed an isolated $^{19}\text{F}-^{13}\text{C}$ two-spin system with a dipole--dipole coupling constant of $b_{^{19}\text{F}^{13}\text{C}}/(2\pi) = -12.1$ kHz. The initial operator was set to $x$-phase $^{19}$F coherence ($I_x$) and detection operator was set to $x$-phase $^{13}$C coherence ($S_x$). 
Powder averaging was in this case  performed using REPULSION\cite{REPULSION} adapted to the size of the chemical shielding anisotropy. For $\omega^\text{aniso}/(2\pi)$ in the ranges of 0-50, 50-100, 100-150 and >150 kHz, 100/144/256/320 $(\alpha_\text{CR}, \beta_\text{CR})$ angles and 10/12/16/18 $\gamma_\text{CR}$ angles were used, respectively. CSA and offset were applied only on the $^{19}$F spin, with an asymmetry parameter $\eta = 0.5$. The mixing time for each experiment was optimized at zero offset and $\omega^\text{aniso}_{^{19}\text{F}}/(2\pi)$ = 80 kHz ($\omega^\text{aniso}_{^{19}\text{F}}/(2\pi)$ = 0 kHz for CP) and subsequently kept constant across the full parameter space.

All simulations assumed a magic-angle spinning frequency of $\omega_r/(2\pi)$ = 25 kHz.  The RF field strengths and additional simulation parameters are provided in the Supplementary  Information and in the scripts available online.

\subsection{Experimental Settings}
All experiments were carried out on a 16.4 T (700 MHz for $^1$H) Bruker Avance III HD wide-bore NMR spectrometer equipped with a 2.5 mm HXY triple-resonance MAS probe, using a powder sample of octafluoronaphthalene (OFN) at ambient temperature. The experiments involved $^{19}$F $\rightarrow$ $^{13}$C  cross polarization with the proposed QOpt pulse sequences compared with standard cross-polarization MAS (CP/MAS),\cite{CPMAS,DCP} ramped CP/MAS (ramp on the $^{13}$C RF channel),\cite{METZ1994219} and $^\text{RESPIRATION}$CP.\cite{jain2012HetCPRESPIRATION,adRESPIRATION,BB-RESPIRATION-CP,nielsen2016theoreticalRESPIRATION} All experiments were performed at a MAS frequency of $\omega_r/(2\pi)$ = 25 kHz.  Cross-polarization data  were recorded using a recycle delay of 4 s and 64 transients. 
For reference, single-pulse MAS experiments were recorded for $^{13}$C and $^{19}$F (150 s recycle delay, 32 transients).

The $^{19}$F $\rightarrow$ $^{13}$C coherence transfer efficiency  was extracted as $\epsilon_\text{FC}=(\gamma_\text{C}/\gamma_\text{F})(\int S_\text{FC})/(\int S_\text{C})$ 
where $\int S_\text{FC}$ is the integrated $^{13}$C signal after transfer and $\int S_\text{C}$ is the corresponding integral from the single-pulse $^{13}$C experiment.
All datasets were normalized to the transfer efficiency at zero offset. 
Pulse sequence details and further experimental parameters are provided in the Supplementary  information.

\section{Results and Discussion}

To demonstrate the capability of SSV-EHT single-spin quaternion optimal control, we address transfer of $^{19}$F coherence to $^{13}$C coherence under the influence of a dominant CSA and a broad span of offsets  on the transferring spin, here specifically $\omega_{^{19}\text{F}}^\text{aniso}/(2\pi)$ = 80 kHz (corresponding to 120 ppm for 700 MHz ($^1$H) instrumentation), and an isotropic chemical shift offset range of 70 kHz (106 ppm in the present case).   Optimizations were conducted as a single-spin pulse optimizations for the RF channel (here $^{19}$F) influenced by large anisotropic shielding.

To ensure recoupling over a large span of isotropic and anisotropic chemical shift values, it is important that the pulse sequence element  generates a non-zero stable effective field over the desired bandwidth as demonstrated previously in the context of DNP pulse sequence development,\cite{PLATO_adv,rCW_EEHT_DNP} and of heteronuclear polarization transfer under MAS\cite{DNP_SSNMR_Carvalho:2025aa, baligacs2026Accordion}. In the present optimization, effective-field mediated rotations corresponding to nominal flip angles of $90^\circ$, $180^\circ$, $270^\circ$, or $360^\circ$ were deliberately avoided  to reduce the risk of matching unfavorable RF-MAS-induced resonances during the modulation period. This argues for our  selected target rotation of $120^\circ$
about the $x$ axis. The modulation time, $\tau_m$, was chosen as $80~\mu$s represented by $n=40$ piecewise-constant pulse elements, yielding a time step of $\Delta t = 2~\mu$s. The selected target rotation of 120$^\circ$ and modulation time of $\tau_m$ = 80 \textmu s correspond to a desired effective field strength of $\omega_\text{eff}/(2\pi)$ = 4.17 kHz.
This choice reflects three considerations. First, long modulation times may lead to overlapping ZQ and DQ recoupling conditions,\cite{DNP_SSNMR_Carvalho:2025aa,rCW_EEHT_DNP} which become more pronounced for strong dipole-dipole coupling interactions, as in our chosen sample object with directly bonded $^{13}$C and $^{19}$F spins in octafluoronaphthalene (OFN). Second, for such relatively strongly coupled nuclei, shorter buildup times are expected, making shorter modulation times advantageous as it allows for precise adjustment of the optimal coherence transfer mixing time for maximum transfer efficiency.\cite{baligacs2026Accordion} Since $\tau_m$ needs to be synchronized with the rotor period, we used a sample spinning frequency of $\omega_r/(2\pi)$ = 25 ~kHz MAS corresponding to a rotor period of $\tau_r$ = 40~$\mu$s. Third, the effective field should be strong enough to ensure effective rotations in the chosen ZQ or DQ operator space (second-averaging of undesired Hamiltonian components) while being weak to avoid excessive influence from RF inhomogeneity.\cite{rCW_EEHT_DNP}  

\begin{figure*}
    \centering
    \includegraphics[width=\columnwidth]{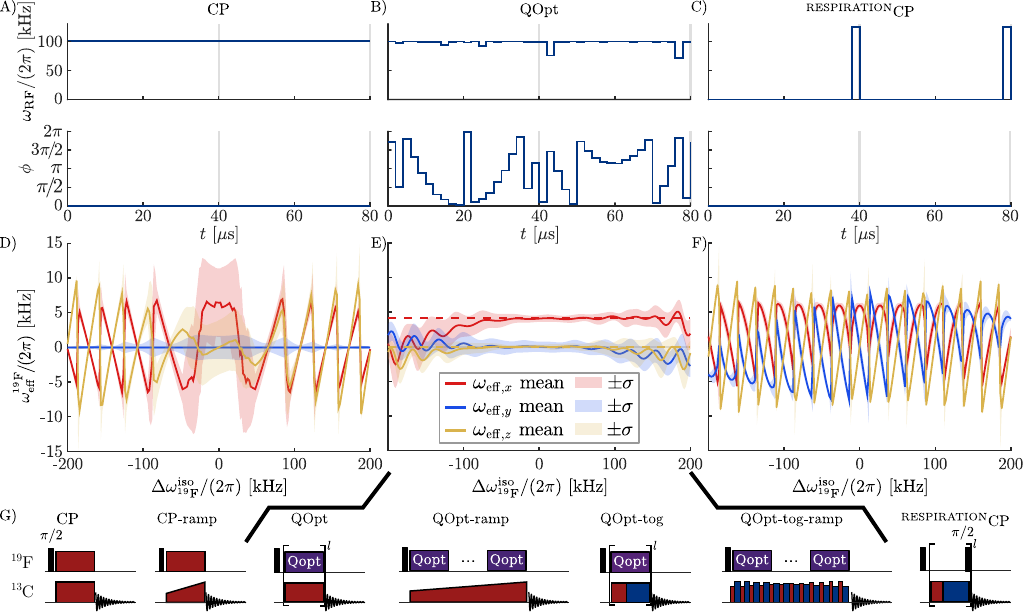}
    \caption{
    \footnotesize
    Comparison of single-spin (I = $^{19}$F) pulse sequence elements, their effective-field properties, and the corresponding two-spin (IS) recoupling experiments along with reference experiments.
    A)–C) $^{19}$F single-spin pulse elements: A) CP (square pulse, $\omega_\text{RF}^\text{I}/(2\pi)$ = 100 kHz), B) QOpt pulse (maximum RF amplitude 100 kHz), and C) rotor-synchronized  2 µs $\pi/2$ pulses used on the broadband-channel in $^\text{RESPIRATION}$CP ($\omega_\text{RF}^\text{I}/(2\pi)$ = 125 kHz). For each pulse sequence, the upper panel shows the RF amplitude and the lower panel the phase (in radians) as a function of time. The total modulation time is $\tau_m = 80~\mu$s, corresponding to 2 rotor periods at 25 kHz MAS.
    D)–F) Effective-field analysis of the pulse elements shown in A)–C) under conditions of $\omega^\text{aniso}_{^{19}\text{F}}/(2\pi)$ = 80 kHz. The Cartesian components of the effective field, $\omega_\text{eff}^{^{19}\text{F}}/(2\pi)$, are shown as a function of $^{19}$F isotropic chemical shift offset, $\Delta\omega_{^{19}\text{F}}^\text{iso}/(2\pi)$. Solid lines represent the mean effective field averaged over 300 uniformly distributed  crystallite orientations (10 $\gamma_{PR}$ and 30 REPULSION $\alpha_{PR}$,$\beta_{PR}$ pairs of angles) while shaded regions represent the  one-standard-deviation interval (i.e., $\pm \sigma$, containing 68\% of all crystallite contributions to the effective field). In E), dotted lines indicate the target effective-field values used during optimization (i.e., 4.17 kHz along $I_x$; 0 kHz along $I_y$ and $I_z$.
    G) Schematic representation of  7 two-spin pulse sequences based on  I-spin ($^{19}$F) single-spin elements: CP and CP-ramp (left), QOpt-based sequences (center), and $^\text{RESPIRATION}$CP
    (right). The S-spin channel ($^{13}$C)  channel is formed by  matching with a simple square pulse, a linearly ramped pulse, '-ramp', or an asymmetrical $x$/$-x$ phase toggled pulse, '-tog', or  a mixture of these where the amplitude of each $x$-phase and $-x$-phase pulse is increased/decreased, respectively.  Solid lines link the single-spin pulse elements in A)–C) give corresponds to the two-spin pulse sequences.
    }
    \label{fig:fig1}
\end{figure*}

The RF amplitudes and phases for the optimized QOpt single-spin pulse element are shown in Fig.~\ref{fig:fig1}B flanked by two reference pulse sequence elements: a simple square pulse corresponding to conventional cross polarization (CP) in Fig.~\ref{fig:fig1}A , and a rotor-synchronized  $\pi/2$-pulse-delay element here chosen as representative for the 'broadband' channel (i.e., the channel without spin lock) in $^\text{RESPIRATION}$CP \cite{jain2012HetCPRESPIRATION} in Fig.~\ref{fig:fig1}C. To assess their suitability for recoupling under different offsets under the condition of 80 kHz anisotropic shielding, all 3 pulse elements were analysed at the single-spin level in terms of their effective Hamiltonians. This analysis used averaging of contributions from 300 uniformly distributed crystallite orientations, and an offset range of $400$~kHz centered around resonance and sampled as 201 equally spaced offset points. The corresponding results are shown in Figs.~\ref{fig:fig1}D--F. 

To ensure cross-polarization over a wide range of I-spin ($^{19}$F) offsets upon matching with an appropriate pulse sequence on the S-spin ($^{13}$C), a desirable optimized I-spin pulse sequence element should provide an effective field 
that remains stable over the desired broad range of offsets and thereby exhibits a narrow span of variation of the effective Hamiltonian upon averaging over anisotropic shielding contributions from crystallites within a uniformly distribution of powder  angles. In Figs.~\ref{fig:fig1}D--\ref{fig:fig1}F, the colored lines represent the crystallite-averaged effective-field components, $x$, $y$, and $z$, while the shaded areas indicate the corresponding one-standard-deviation interval ($\pm \sigma$). This interval   includes $\approx 68$ \% of the crystal-angle contributions. The target values for the $x$-phase effective field used in the QOpt optimization are marked by dotted lines in Fig.~\ref{fig:fig1}E. The pulse sequence was optimized for an $^{19}$F isotropic chemical shift bandwidth of 80 kHz. Among the three pulse sequences analyzed, the optimized QOpt pulse element clearly provides the most stable effective field over the desired bandwidth of $-40$ to $40$~kHz and well beyond that.  A stable effective field is achieved for a span of isotropic chemical shift offsets approaching $200$~kHz. In contrast, the square pulse element of the conventional CP experiment produces a much larger distribution of effective fields for a powder  than our optimization target, and is markedly less stable with respect to offset. The rotor-synchronized train of $\pi/2$  pulses of the $^\text{RESPIRATION}$CP  experiment exhibit a very narrow crystallite spread, but show the characteristic periodic and narrow offset dependence imposed by the rotor-synchronized RF modulation. This conforms well with expectations.\cite{jain2012HetCPRESPIRATION,adRESPIRATION,BB-RESPIRATION-CP,nielsen2016theoreticalRESPIRATION} We note that  $^\text{RESPIRATION}$CP is included in the comparison as it provides, within this very narrow isotropic chemical shift range,  extremely good compensation towards anisotropic chemical shielding effects here achieved with much less accumulated RF power than the other sequences. The challenge is the narrow chemical shift range which complicates measurement on samples with a larger offset span. 

The three explored $^{19}$F  pulse sequence elements may be combined with a large variety of matching $^{13}$C pulse sequence elements to realize heteronuclear coherence transfer within a two-spin system. Seven such combined pulse sequences  are shown in Fig.~\ref{fig:fig1}G. These include, taken from the left, conventional cross polarization (CP) and its ramped variant (CP-ramp). For the polarization transfer with our QOpt element, the element is repeated on the $^{19}$F channel until the desired buildup time is reached. On the second channel, it can be combined with different pulse elements: a simple square pulse (QOpt), a ramped pulse analogous to CP-ramp (QOpt-ramp), a toggled pulse sequence alternating between an $x$-phase pulse and a slightly longer $-x$-phase pulse (QOpt-tog), and a ramped version of this toggled element (QOpt-tog-ramp), which was obtained by gradually increasing the $x$-phase RF amplitude while simultaneously decreasing the $-x$-phase RF amplitude. The toggled schemes are designed to provide a broader and more favorable effective-field profile on the S-spin ($^{13}$C) channel than obtained using a simple  square pulse. Finally, the rotor-synchronized $\pi/2$ pulses can be combined with an phase-alternating pulse train on the second RF channel to yield the $^\text{RESPIRATION}$CP experiment, implemented with different length or amplitude of the $x$ and $-x$ elements.\cite{jain2012HetCPRESPIRATION}

\begin{figure*}
    \centering
    \includegraphics[width=1\linewidth]{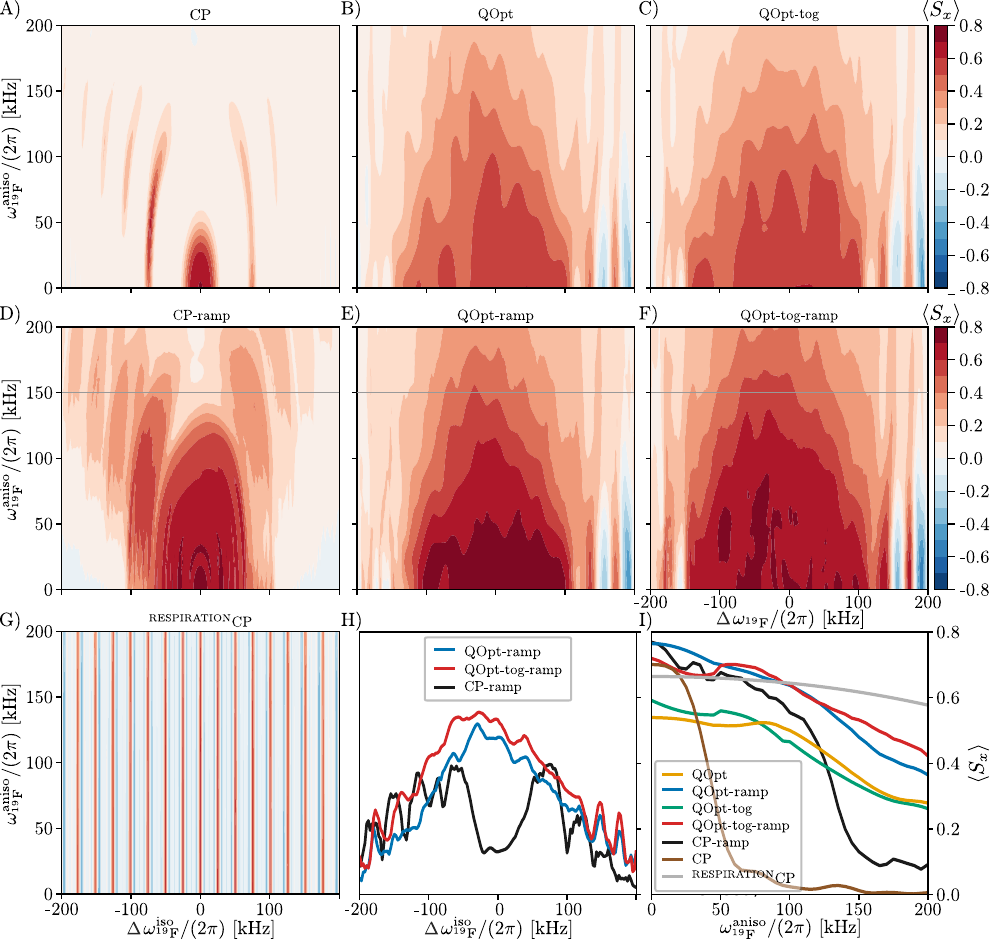}
    \caption{
    \footnotesize
    A-G) Two-spin SIMPSON simulations of $^{19}$F$\rightarrow^{13}$C $x$-phase coherence transfer efficiency as a function of $^{19}$F chemical shift offset ($\Delta\omega_{^{19}\text{F}}^\text{iso}/(2\pi)$) and chemical shielding anisotropy (CSA, $\omega^\text{aniso}_{^{19}\text{F}}/(2\pi)$) for  A) CP, B) QOpt, C) QOpt-tog, D) CP-ramp, E) QOpt-ramp, F) QOpt-tog-ramp, and G) $^\text{RESPIRATION}$CP. The color scale represents the  efficiency of transfer  $\langle S_x \rangle$ with S representing $^{13}$C.
    H) Transfer efficiency as a function of $^{19}$F offset at a fixed CSA of $\omega_{^{19}\mathrm{F}}^\mathrm{aniso}/(2\pi)=150$ kHz for the 3 best-performing pulse sequences (D--F). These traces correspond to the horizontal grey lines in panels D–F and are shown to facilitate a direct comparison of the offset robustness of the 3 best-performing pulse sequences.    
    I) Transfer efficiency as a function of $^{19}$F CSA at zero isotropic offset for all seven pulse sequences, illustrating their relative tolerance toward increasing CSA in the absence of resonance offset.}
    \label{fig:fig2}
\end{figure*}

With optimized (in QOpt) pulse sequences and the pulse sequences chosen for comparison characterized by their single-spin effective-field properties, the next step is to evaluate their performance for coherence transfer in a two-spin setting. Figure~\ref{fig:fig2} shows numerical simulations of the transfer efficiency ($\langle S_x \rangle$) over a grid of $^{19}$F isotropic chemical shift offset $\Delta \omega^\text{iso}_{^{19}\text{F}}/(2\pi)$ taken over values between $-200$ and $200$~kHz and anisotropic chemical shift $\omega_{^{19}\text{F}}^\mathrm{aniso}/(2\pi)$ values between $0$ and $200$~kHz and  for all seven experiments reported in Fig.~\ref{fig:fig1}G. As expected, the ramped variants generally outperform their square-pulse counterparts for CP, QOpt, and QOpt-tog. In the vicinity of a CSA of $70$~kHz, conventional CP-ramp still performs reasonably well, and CP-ramp recovers a substantial part of the transfer efficiency that is lost between the narrow matching regions of standard CP. The ramp therefore extends the addressable offset range and improves performance in case of larger CSA's. Nevertheless, even CP-ramp retains distinct regions in which no efficient transfer is achieved, including within the experimentally relevant regime around $70$~kHz CSA for OFN.

It is clearly evident that the experiments based on the optimized QOpt pulse element display a more homogeneous transfer profile over a broader offset range and remain effective up to larger CSA values. Notably, the QOpt-based scheme already performs well when combined with a simple square pulse on the $^{13}$C (S spin) RF channel, suggesting that the improved robustness originates primarily from the single-spin optimized effective-field behavior of the $^{19}$F (I spin) pulse element itself. Additional improvements may be obtained when the $^{13}$C pulse sequence is replaced by a ramped or toggled element, as clearly evidenced by broadening of the range of efficient transfer. Overall, these simulations demonstrate that the QOpt design strategy yields recoupling schemes that are substantially less sensitive to offsets and CSA than conventional CP-based approaches.

\begin{figure*}
    \centering
    \includegraphics[width=1.0\columnwidth]{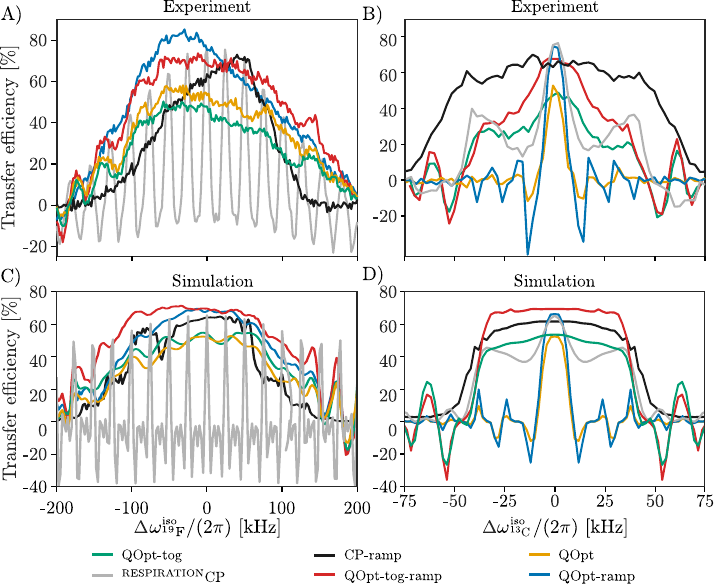}
    \caption{
    \footnotesize
    Experimental (A,B) and numerically simulated (C,D) $^{19}$F$\rightarrow^{13}$C coherence transfer efficiencies for the recoupling experiments shown in Fig.~\ref{fig:fig1}G as a function of offset for OFN (CSA $\approx 70$ kHz) under conditions of $\omega_r/(2\pi)=25$ kHz MAS. Panels (A,C) show the $^{19}$F offset dependence, while panels (B,D) show the $^{13}$C offset dependence. 
    Simulations were performed using the same parameters as in Fig.~\ref{fig:fig2} assuming $\omega^\text{aniso}_{^{19}\text{F}}/(2\pi)$ = 70 kHz. Mixing times were optimized independently for each experiment. The CP experiment is omitted due to negligible transfer efficiency (cf., Fig.~\ref{fig:fig2}A)). 
    }
    \label{fig:fig3}
\end{figure*}

Finally, we compare in Fig.~\ref{fig:fig3} the simulated results with experimental data recorded for OFN using 700 MHz (at $^1$H) instrumentation with 25 kHz MAS. Overall, the experiments reproduce the main features predicted by the numerical simulations in Fig.~\ref{fig:fig2} very well, both in terms of transfer efficiency and offset dependence. 
This confirms that the single-spin effective-field description and experiment design strategy presented in Fig.~\ref{fig:fig1}  provides an accurate framework for designing efficient pulse sequences with predictive experimental recoupling performance.

While the behavior on the $^{19}$F channel is largely identical across the different QOpt-based schemes all performing better than CP-ramp, more significant differences arise on the $^{13}$C channel. As evidenced from Figs.~\ref{fig:fig3}B,\ref{fig:fig3}D, the QOpt and QOpt-ramp experiments exhibit relatively narrow offset profiles on the carbon channel, despite their robust performance on the $^{19}$F channel. Opposedly, the toggled variants (QOpt-tog and QOpt-tog-ramp) show a substantially broader offset range on the $^{13}$C  channel, while leaving the $^{19}$F channel performance essentially unchanged. These results demonstrate that the effective-field properties for the two coupled spin species can largely be controlled  independently: the optimized QOpt pulse defines the behavior on the $^{19}$F channel, whereas the choice of pulse element on the second channel determines the bandwidth of the matching condition.

The improved performance of the toggled schemes can be understood from the effective-field perspective. By alternating between $x$- and $-x$-phase  pulses, the same average effective field is obtained by increasing the RF amplitudes. This leads to a more uniform effective-field magnitude over a broader offset range, thereby extending the matching condition on the second channel. More generally, this reinforces that the second-channel element can be designed independently, and that more sophisticated pulse shapes may be employed as long as the effective-field condition derived in Eq.~(\ref{eqn:HeffCondition}) is satisfied. This opens up the possibility for single-spin based design design of two-spin cross-polarization methods, like in the example in this work, but also for more general optimal control optimization design of pulse sequences with tailored properties beyond broadband offset compensation.

Quantitatively, some deviations between experiment and simulation are observed. CP-ramp shows a somewhat broader offset profile experimentally than predicted numerically. This may originate from slightly higher effective RF amplitudes or differences in the experimentally optimized contact times. Conversely, the toggled sequences appear slightly narrower and with a reduced transfer efficiency away from the on-resonance case in experiment than in simulation. This likely reflects the increased difficulty of experimentally optimizing these sequences compared to the more systematic optimization possible in simulations. Nevertheless, the overall agreement in both width and transfer efficiency is good.

Taken together, these results demonstrate that the effective-field-based design strategy translates directly from simulation to experiment. In particular, the combination of an optimized pulse on the $^{19}$F channel with a flexible, independently tunable element on the $^{13}$C  channel provides a practical route to robust heteronuclear recoupling over a broad range of offsets and CSA values.

\section{Conclusions}

We have presented a quaternion-based optimal control optimization approach for the design of heteronuclear dipolar recoupling pulse elements under MAS in the presence of large chemical shift anisotropy. Exploiting the single-spin-vector effective Hamiltonian formalism, the optimization task is reduced to single-spin RF irradiation invoked via pulse sequence elements with short modulation time, which for cross polarization applications may be repeated to achieve optimal transfer compatible with relevant dipolar coupling interactions. This enables efficient quaternion-based optimal control, which was instantiated as our optimization tool.

Using this framework, we designed $^{19}$F pulse sequence elements that provide a stable effective-field profile over a broad offset range and across many crystallite orientations in the presence of large CSA. The optimal pulse element was compared with two established dipolar recoupling elements in terms of components of their effective Hamiltonian. In addition, detailed SIMPSON simulations were carried out for seven different recoupling schemes, four of which employed the optimized QOpt pulse sequence element. These simulations demonstrated that the optimal QOpt element yields markedly more homogeneous and robust transfer over broad offset and CSA ranges than conventional CP-based approaches. Experimental results obtained for $^{19}$F $\rightarrow$  $^{13}$C on OFN were in good agreement with the simulations and confirmed the predicted improvement in recoupling performance.

Beyond the specific pulse demonstrated here, the results illustrate a more general design principle: once the effective field on one channel has been optimized, the pulse element on the second channel can be adjusted independently, provided that the effective-field matching condition is fulfilled, and can therefore be tailored to specific needs, such as isotropic and anisotropic chemical shielding in the present example. This opens a route not only to improved broadband heteronuclear recoupling under challenging CSA conditions, but also to more flexible pulse sequence design based on the combination of the SSV-EHT framework and numerical optimization.

\section*{Author Contributions}
All authors contributed to the design of the research, supervised by A.B.N. and N.C.N.. E.B. and A.B.N. performed all experiments. E.B., A.B.N., and N.W. developed the optimizer tool. E.B. did numerical calculations supported by A.B.N. and J.P.C., E.B, and N.C.N. contributed to the initial draft  with all authors involved in the final manuscript. N.C.N. ensured funding for the project. 

\section*{Conflicts of interest}
There are no conflicts to declare.

\section*{Acknowledgements}
We acknowledge financial support from The  Independent Research Fund Denmark (Grant 2032-00215B), the Villum Foundation (Grant 50099), the Novo Nordisk Foundation (Grant NNF22OC0076002), and the DeiC National HPC (g.a. DeiC-AU-N5-2025153).

\bibliography{rsc} 
\bibliographystyle{sciencemag}


\subsection*{Supplementary materials}
\sloppy
S1. Quaternion Formalism\\
S2. Analytical Gradient\\
S3. Chemical Shielding Anisotropy Fitting\\
S4. Materials and methods\\
S5. QOpt Pulse Sequence\\


\newpage


\renewcommand{\thefigure}{S\arabic{figure}}
\renewcommand{\thetable}{S\arabic{table}}
\renewcommand{\theequation}{S\arabic{equation}}
\renewcommand{\thepage}{S\arabic{page}}
\setcounter{figure}{0}
\setcounter{table}{0}
\setcounter{equation}{0}
\setcounter{page}{1} 

\clearpage

\appendix
\setcounter{section}{0}
\setcounter{figure}{0}
\setcounter{table}{0}

\renewcommand{\thesection}{S\arabic{section}}

\renewcommand{\thefigure}{S\arabic{figure}}

\renewcommand{\thetable}{S\arabic{table}}

\begin{center}
\section*{Supplementary Materials for\\ \scititle}

	Enik\H{o} Baligács, Nino Wili, José P. Carvalho,
	Anders Bodholt Nielsen,$^{\ast}$
	  and
    Niels Chr. Nielsen$^{\ast}$
 \and
	\small Interdisciplinary Nanoscience Center (iNANO) and Department of Chemistry, Aarhus University, 
    \\ \small Gustav Wieds Vej 14, DK-8000 Aarhus C, Denmark.
    \\
	\small$^\ast$Corresponding authors. Email: ncn@chem.au.dk and abn@chem.au.dk \and

\end{center}


\subsubsection*{This PDF file includes:}
\sloppy
S1. Quaternion Formalism\\
S2. Analytical Gradient\\
S3. Chemical Shielding Anisotropy Fitting\\
S4. Materials and methods\\
S5. QOpt Pulse Sequence\\

\newpage

\vspace{3em}

\clearpage
\section{Quaternion Formalism }

Throughout this work, quaternions are represented in scalar-first form,
\begin{equation}
    \vec q =
    \begin{bmatrix}
    A\\ B\\ C\\ D
    \end{bmatrix},
    \qquad
    q=A+iB+jC+kD,
\end{equation}
where $A$ is the scalar part and $(B,C,D)^T$ denotes the vector part. The quaternion units satisfy
\begin{equation}
    i^2=j^2=k^2=ijk=-1,
\end{equation}
which implies
\begin{equation}
    ij=k,\quad jk=i,\quad ki=j,\qquad ji=-k,\quad kj=-i,\quad ik=-j.
\end{equation}

For two quaternions $q_2=A_2+iB_2+jC_2+kD_2$ and $q_1=A_1+iB_1+jC_1+kD_1$, their product is obtained by expansion and collection of scalar, $i$, $j$, and $k$ terms,
\begin{align}
q_2q_1
=& (A_2+iB_2+jC_2+kD_2)(A_1+iB_1+jC_1+kD_1) \nonumber\\
=& ~~~~~A_2A_1 + ~A_2iB_1 + ~A_2jC_1 + ~A_2kD_1 \\
& + iB_2A_1 + iB_2iB_1 + iB_2jC_1 + iB_2kD_1 \nonumber\\
&+ jC_2A_1 + jC_2iB_1 + jC_2jC_1 + jC_2kD_1 \\
&+ kD_2A_1 + kD_2iB_1 + kD_2jC_1 + kD_2kD_1 \nonumber\\
=& A_2A_1 + iA_2B_1 + jA_2C_1 + kA_2D_1 \\
&+ iB_2A_1 - B_2B_1 + kB_2C_1 - jB_2D_1 \nonumber\\
&+ jC_2A_1 - kC_2B_1 - C_2C_1 + iC_2D_1 \\
& + kD_2A_1 + jD_2B_1 - iD_2C_1 - D_2D_1 \nonumber\\
=& A_2A_1 - B_2B_1 - C_2C_1 - D_2D_1 \\
& + i(A_2B_1+B_2A_1+C_2D_1-D_2C_1) \nonumber\\
&+ j(A_2C_1-B_2D_1+C_2A_1+D_2B_1) \\
& + k(A_2D_1+B_2C_1-C_2B_1+D_2A_1).
\end{align}

Thus, in four-component vector form,
\begin{equation}
\vec q_2\otimes\vec q_1 =
\begin{bmatrix}
A_2A_1-B_2B_1-C_2C_1-D_2D_1\\
A_2B_1+B_2A_1+C_2D_1-D_2C_1\\
A_2C_1-B_2D_1+C_2A_1+D_2B_1\\
A_2D_1+B_2C_1-C_2B_1+D_2A_1
\end{bmatrix}.
\end{equation}

Equivalently, the same operation may be written as
\begin{equation}
\vec q_2\otimes\vec q_1 =
\begin{bmatrix}
 A_2 & -B_2 & -C_2 & -D_2\\
 B_2 &  A_2 & -D_2 &  C_2\\
 C_2 &  D_2 &  A_2 & -B_2\\
 D_2 & -C_2 &  B_2 &  A_2
\end{bmatrix}
\begin{bmatrix}
A_1\\ B_1\\ C_1\\ D_1
\end{bmatrix}.
\end{equation}

\section{Analytical Gradient}

The analytical gradients required for optimization are obtained by differentiation of the corresponding quaternion with respect to the optimization parameters, namely the RF amplitude $\omega_{\mathrm{RF},j}$ and RF phase $\phi_{\mathrm{RF},j}$. The derivative with respect to the RF amplitude is

\begin{equation}
\frac{\partial \vec q_{j,\Omega}}{\partial \omega_{\mathrm{RF},j}}=\begin{bmatrix}-\frac{\Delta t_j}{2}\frac{\omega_{\mathrm{RF},j}}{|\omega_{j,\Omega}|}\sin\theta_{j,\Omega}\\ -\left(\frac{\Delta\omega_\Omega^2}{|\omega_{j,\Omega}|^3}\sin\theta_{j,\Omega}+\frac{\omega_{\mathrm{RF},j}^2\Delta t_j}{2|\omega_{j,\Omega}|^2}\cos\theta_{j,\Omega}\right)\cos\phi_{\mathrm{RF},j}\\ -\left(\frac{\Delta\omega_\Omega^2}{|\omega_{j,\Omega}|^3}\sin\theta_{j,\Omega}+\frac{\omega_{\mathrm{RF},j}^2\Delta t_j}{2|\omega_{j,\Omega}|^2}\cos\theta_{j,\Omega}\right)\sin\phi_{\mathrm{RF},j}\\ \Delta\omega_\Omega\left(\frac{\omega_{\mathrm{RF},j}}{|\omega_{j,\Omega}|^3}\sin\theta_{j,\Omega}-\frac{\omega_{\mathrm{RF},j}\Delta t_j}{2|\omega_{j,\Omega}|^2}\cos\theta_{j,\Omega}\right)\end{bmatrix}.
\end{equation}
where $|\omega_{j,\Omega}|=\sqrt{\omega_{\mathrm{RF},j}^{2}+\Delta\omega_{\Omega}^{2}}$. 

For the RF phase, only the two transverse quaternion components depend explicitly on $\phi_{\mathrm{RF},j}$. The derivative is therefore
\begin{equation}
\frac{\partial \vec q_{j,\Omega}}{\partial \phi_{\mathrm{RF},j}}=\begin{bmatrix}0\\ \sin\!\left(\frac{|\omega_{j,\Omega}|\Delta t_j}{2}\right)\frac{\omega_{\mathrm{RF},j}}{|\omega_{j,\Omega}|}\sin\phi_{\mathrm{RF},j}\\ -\sin\!\left(\frac{|\omega_{j,\Omega}|\Delta t_j}{2}\right)\frac{\omega_{\mathrm{RF},j}}{|\omega_{j,\Omega}|}\cos\phi_{\mathrm{RF},j}\\ 0\end{bmatrix}.
\end{equation}

\clearpage

\section{Chemical Shielding Anisotropy Fitting}
To determine the $^{19}$F CSA ($\omega_{^{19}F}^\text{aniso}$)), several reference 1D spectra were recorded on  700 MHz and  400 MHz instrumentation. These spectra were compared to SIMPSON simulations to determine the CSA. 

The fittings around the determined region is shown below: 
\begin{figure}
    \centering
    \includegraphics[width=\textwidth]{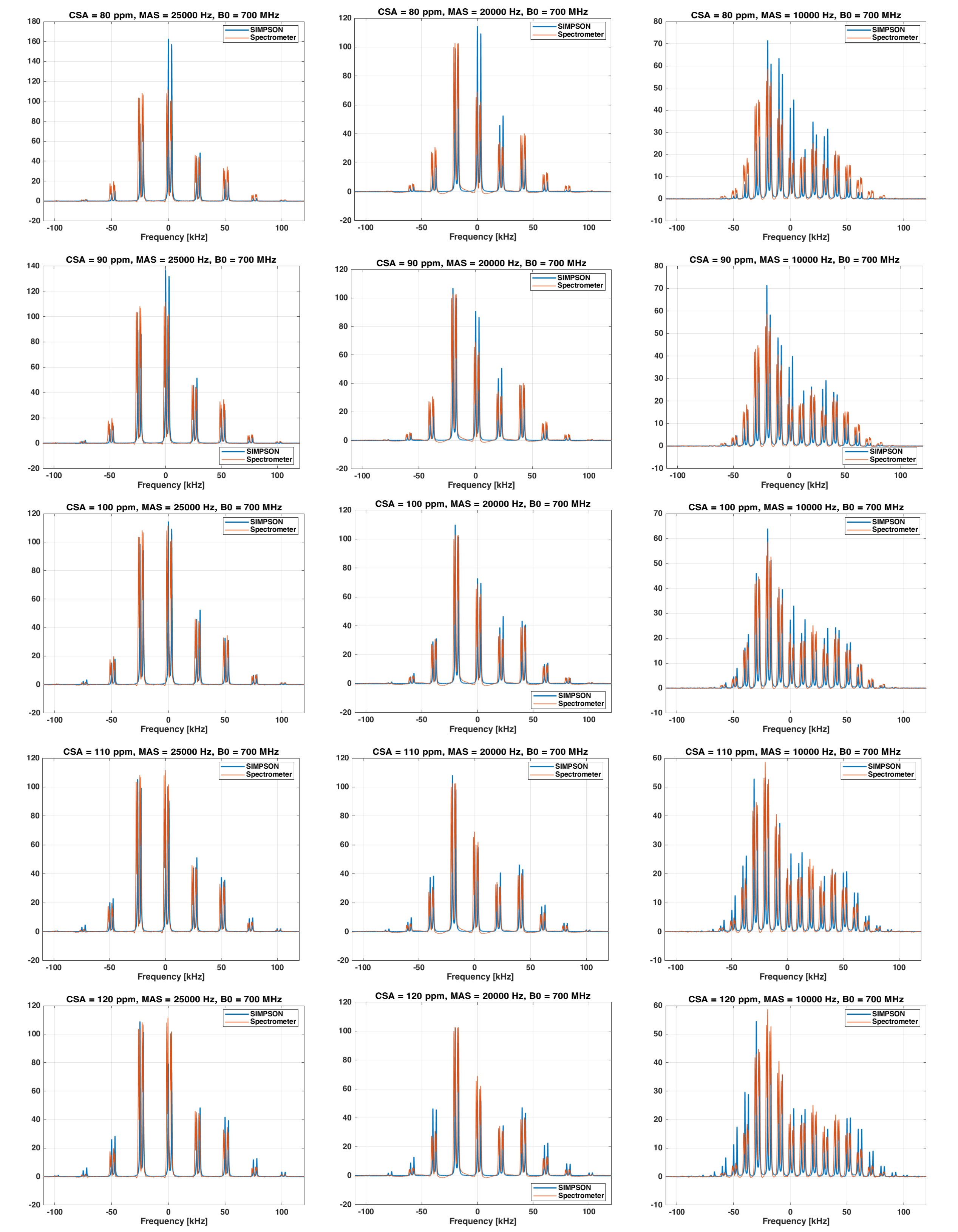}
    \caption{\footnotesize Fittings at a magnetic field at 700 MHz}
    \label{fig:CSAfitting700}
\end{figure}

\begin{figure}
    \centering
    \includegraphics[width = 0.7\textwidth]{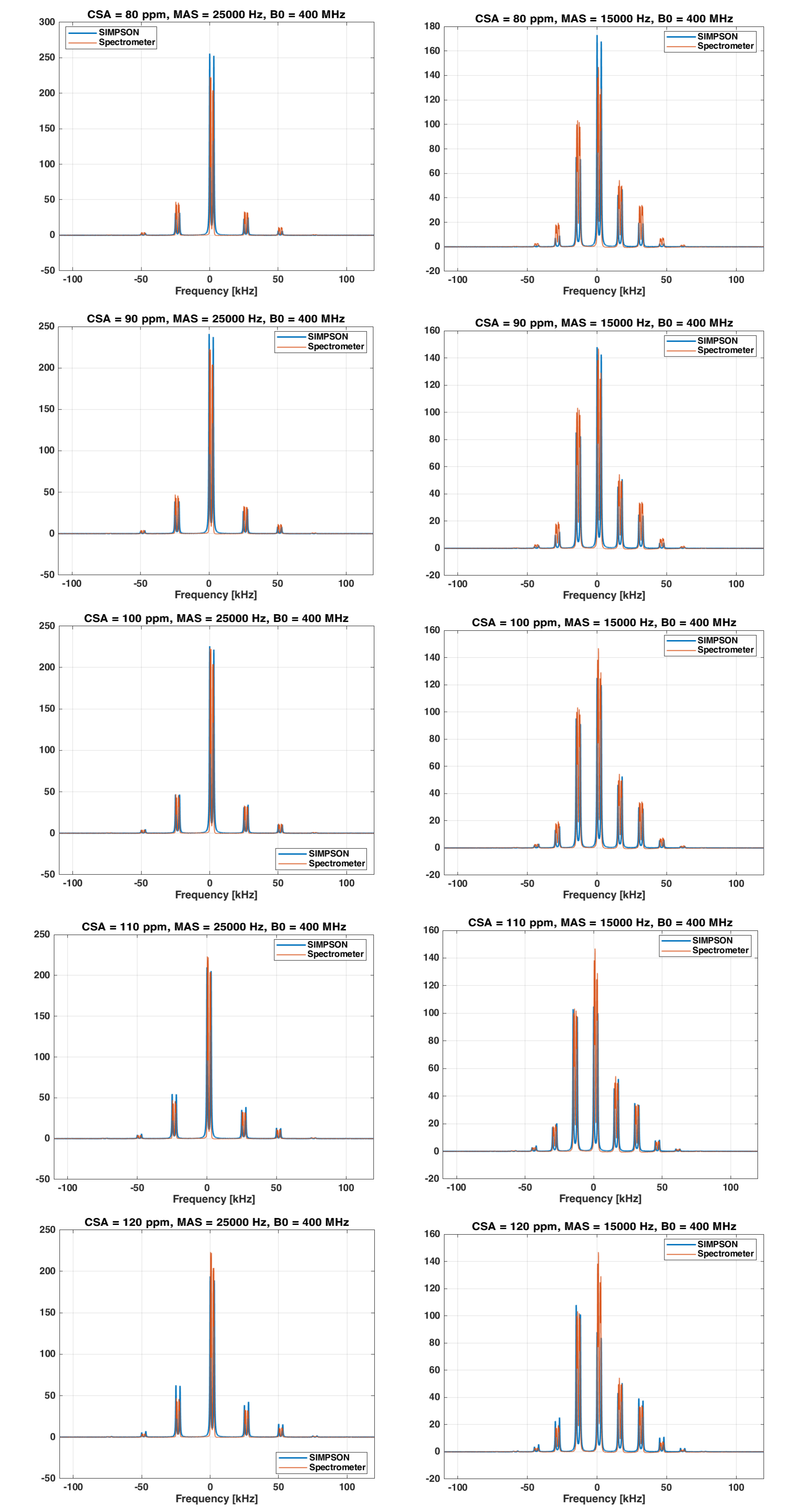}
    \caption{\footnotesize Fittings at a magnetic field at 400 MHz}
    \label{fig:CSAfitting400}
\end{figure}

\clearpage
\section{Materials and methods}

Additional simulation and experimental parameters not given in the main text are summarized in Tables~\ref{tab:SimParameters} and \ref{tab:ExpParameters}.

For each pulse sequence, the RF field strengths, optimized contact times ($\tau_\mathrm{mix}$), and additional sequence-specific parameters were optimized individually by maximizing the polarization transfer efficiency and are listed in Table~\ref{tab:ExpParameters}. For pulse sequences employing repeated pulse blocks, the corresponding loop number $l$ is also given, where the total contact time is related to the pulse block duration by $\tau_\mathrm{mix}=l\tau_m$. For experiments employing RF ramps, linear amplitude ramps were used, with the corresponding ramp ranges listed in the final column.

\begin{table}[ht]
\centering
\scriptsize
\begin{tabular}{lccccc}
\hline
Experiment &
$\omega_{\mathrm{RF}}^{^{19}\mathrm{F}}/(2\pi)$ [kHz] &
$\omega_{\mathrm{RF}}^{^{13}\mathrm{C}}/(2\pi)$ [kHz] &
$\tau_{\mathrm{mix}}$ [µs] &
$l$ &
Ramp [\%] \\
\hline
CP              & 100.0 & 75.0 &  760 & -  &  - \\
CP-ramp         & 100.0 & 75.0 &  722 & -  & 70-100 \\
QOpt            & 100.0 & 16.5 &  400 &  5  &  - \\
QOpt-ramp       & 100.0 & 16.5 &  720 &  9 & 80-100 \\
QOpt-tog        & 100.0 & 41.8 &  400 &  5 & -  \\
QOpt-tog-ramp   & 100.0 & 41.8 & 2880 & 36 & 95-100 \\
RESPIRATION     &  62.5 & 40.0 &  200 &  5 &  - \\
\hline
\end{tabular}
\caption{\footnotesize SIMPSON simulation parameters for the polarization-transfer experiments shown in Fig.~2 and Fig.~3.}
\label{tab:SimParameters}
\end{table}

\begin{table}[ht]
\centering
\scriptsize
\begin{tabular}{lccccc}
\hline
Experiment &
$\omega_{\mathrm{RF}}^{^{19}\mathrm{F}}/(2\pi)$ [kHz] &
$\omega_{\mathrm{RF}}^{^{13}\mathrm{C}}/(2\pi)$ [kHz] &
$\tau_{\mathrm{mix}}$ [µs] &
$l$ &
Ramp [\%] \\
\hline
CP-ramp         & 77.9 & 100.0 & 1100 & - & 70-100  \\
QOpt            & 77.9 & 11.7 &  320 &  4 &  - \\
QOpt-ramp       & 80.0 & 14.6 & 2080 & 26 & 50-100 \\
QOpt-tog        & 77.9 & 51.5 &  320 &  4 & -  \\
QOpt-tog-ramp   & 77.9 & 65.3 & 1120 & 14 & See listing \ref{lst:XiX} in section \ref{sec:secv}\\
RESPIRATION     & 83.3 & 40.0 &  280 &  6 & - \\
\hline
\end{tabular}
\caption{\footnotesize Experimental parameters for the polarization-transfer experiments shown in Fig.~3.}
\label{tab:ExpParameters}
\end{table}

\clearpage
\section{QOpt Pulse Sequence}
\label{sec:secv}
Below is shown the QOpt pulse sequence, and the tog-ramp sequence used in this work presented.  The first column represents the pulse amplitude in Hz, the second column represents the phase in degrees, where 0 ° corresponds to an $x$ phase.  Each pulse has a length of 2 µs. 
\begin{lstlisting}[caption={\footnotesize QOpt shape file.}]
99882.6002 	 308.58 	
97593.2747 	 91.90 	
99938.7784 	 322.74 	
99788.5816 	 235.95 	
98277.6433 	 158.68 	
99577.6807 	 107.58 	
98881.0528 	 55.28 	
94404.7994 	 30.57 	
99306.0922 	 8.96 	
99774.6973 	 6.07 	
98040.4008 	 358.64 	
99810.2557 	 19.22 	
91823.9307 	 35.18 	
99879.0133 	 64.16 	
99432.6134 	 115.58 	
99913.2100 	 171.43 	
99930.7510 	 249.82 	
99937.6518 	 334.75 	
99206.1576 	 84.39 	
99416.0881 	 205.60 	
99192.2675 	 18.55 	
75896.7021 	 262.57 	
99512.2651 	 169.77 	
99014.8033 	 79.16 	
99886.4254 	 9.71 	
99490.0204 	 311.73 	
99904.9940 	 264.74 	
99625.5643 	 231.86 	
99500.3110 	 212.56 	
99651.1342 	 208.43 	
99852.4588 	 204.51 	
99754.5915 	 222.47 	
99205.2418 	 242.32 	
99857.1682 	 268.75 	
99945.0147 	 319.18 	
99554.7653 	 13.80 	
99728.1894 	 94.72 	
99434.0476 	 193.81 	
71090.5744 	 327.26 	
98881.1991 	 38.66 	
\end{lstlisting}

\begin{lstlisting}[caption={\footnotesize Tog-ramp pulse sequence file},label={lst:XiX}]
7.70000000E+01, 0.00000000E+00
7.70000000E+01, 0.00000000E+00
7.70000000E+01, 0.00000000E+00
7.70000000E+01, 0.00000000E+00
8.30000000E+01, 1.80000000E+02
8.30000000E+01, 1.80000000E+02
8.30000000E+01, 1.80000000E+02
8.30000000E+01, 1.80000000E+02
8.30000000E+01, 1.80000000E+02
8.30000000E+01, 1.80000000E+02
7.70000000E+01, 0.00000000E+00
7.70000000E+01, 0.00000000E+00
7.70000000E+01, 0.00000000E+00
7.70000000E+01, 0.00000000E+00
8.30000000E+01, 1.80000000E+02
8.30000000E+01, 1.80000000E+02
8.30000000E+01, 1.80000000E+02
8.30000000E+01, 1.80000000E+02
8.30000000E+01, 1.80000000E+02
8.30000000E+01, 1.80000000E+02

7.75000000E+01, 0.00000000E+00
7.75000000E+01, 0.00000000E+00
7.75000000E+01, 0.00000000E+00
7.75000000E+01, 0.00000000E+00
8.25000000E+01, 1.80000000E+02
8.25000000E+01, 1.80000000E+02
8.25000000E+01, 1.80000000E+02
8.25000000E+01, 1.80000000E+02
8.25000000E+01, 1.80000000E+02
8.25000000E+01, 1.80000000E+02
7.75000000E+01, 0.00000000E+00
7.75000000E+01, 0.00000000E+00
7.75000000E+01, 0.00000000E+00
7.75000000E+01, 0.00000000E+00
8.25000000E+01, 1.80000000E+02
8.25000000E+01, 1.80000000E+02
8.25000000E+01, 1.80000000E+02
8.25000000E+01, 1.80000000E+02
8.25000000E+01, 1.80000000E+02
8.25000000E+01, 1.80000000E+02

7.80000000E+01, 0.00000000E+00
7.80000000E+01, 0.00000000E+00
7.80000000E+01, 0.00000000E+00
7.80000000E+01, 0.00000000E+00
8.20000000E+01, 1.80000000E+02
8.20000000E+01, 1.80000000E+02
8.20000000E+01, 1.80000000E+02
8.20000000E+01, 1.80000000E+02
8.20000000E+01, 1.80000000E+02
8.20000000E+01, 1.80000000E+02
7.80000000E+01, 0.00000000E+00
7.80000000E+01, 0.00000000E+00
7.80000000E+01, 0.00000000E+00
7.80000000E+01, 0.00000000E+00
8.20000000E+01, 1.80000000E+02
8.20000000E+01, 1.80000000E+02
8.20000000E+01, 1.80000000E+02
8.20000000E+01, 1.80000000E+02
8.20000000E+01, 1.80000000E+02
8.20000000E+01, 1.80000000E+02

7.85000000E+01, 0.00000000E+00
7.85000000E+01, 0.00000000E+00
7.85000000E+01, 0.00000000E+00
7.85000000E+01, 0.00000000E+00
8.15000000E+01, 1.80000000E+02
8.15000000E+01, 1.80000000E+02
8.15000000E+01, 1.80000000E+02
8.15000000E+01, 1.80000000E+02
8.15000000E+01, 1.80000000E+02
8.15000000E+01, 1.80000000E+02
7.85000000E+01, 0.00000000E+00
7.85000000E+01, 0.00000000E+00
7.85000000E+01, 0.00000000E+00
7.85000000E+01, 0.00000000E+00
8.15000000E+01, 1.80000000E+02
8.15000000E+01, 1.80000000E+02
8.15000000E+01, 1.80000000E+02
8.15000000E+01, 1.80000000E+02
8.15000000E+01, 1.80000000E+02
8.15000000E+01, 1.80000000E+02

7.90000000E+01, 0.00000000E+00
7.90000000E+01, 0.00000000E+00
7.90000000E+01, 0.00000000E+00
7.90000000E+01, 0.00000000E+00
8.10000000E+01, 1.80000000E+02
8.10000000E+01, 1.80000000E+02
8.10000000E+01, 1.80000000E+02
8.10000000E+01, 1.80000000E+02
8.10000000E+01, 1.80000000E+02
8.10000000E+01, 1.80000000E+02
7.90000000E+01, 0.00000000E+00
7.90000000E+01, 0.00000000E+00
7.90000000E+01, 0.00000000E+00
7.90000000E+01, 0.00000000E+00
8.10000000E+01, 1.80000000E+02
8.10000000E+01, 1.80000000E+02
8.10000000E+01, 1.80000000E+02
8.10000000E+01, 1.80000000E+02
8.10000000E+01, 1.80000000E+02
8.10000000E+01, 1.80000000E+02

7.95000000E+01, 0.00000000E+00
7.95000000E+01, 0.00000000E+00
7.95000000E+01, 0.00000000E+00
7.95000000E+01, 0.00000000E+00
8.05000000E+01, 1.80000000E+02
8.05000000E+01, 1.80000000E+02
8.05000000E+01, 1.80000000E+02
8.05000000E+01, 1.80000000E+02
8.05000000E+01, 1.80000000E+02
8.05000000E+01, 1.80000000E+02
7.95000000E+01, 0.00000000E+00
7.95000000E+01, 0.00000000E+00
7.95000000E+01, 0.00000000E+00
7.95000000E+01, 0.00000000E+00
8.05000000E+01, 1.80000000E+02
8.05000000E+01, 1.80000000E+02
8.05000000E+01, 1.80000000E+02
8.05000000E+01, 1.80000000E+02
8.05000000E+01, 1.80000000E+02
8.05000000E+01, 1.80000000E+02

8.00000000E+01, 0.00000000E+00
8.00000000E+01, 0.00000000E+00
8.00000000E+01, 0.00000000E+00
8.00000000E+01, 0.00000000E+00
8.00000000E+01, 1.80000000E+02
8.00000000E+01, 1.80000000E+02
8.00000000E+01, 1.80000000E+02
8.00000000E+01, 1.80000000E+02
8.00000000E+01, 1.80000000E+02
8.00000000E+01, 1.80000000E+02
8.00000000E+01, 0.00000000E+00
8.00000000E+01, 0.00000000E+00
8.00000000E+01, 0.00000000E+00
8.00000000E+01, 0.00000000E+00
8.00000000E+01, 1.80000000E+02
8.00000000E+01, 1.80000000E+02
8.00000000E+01, 1.80000000E+02
8.00000000E+01, 1.80000000E+02
8.00000000E+01, 1.80000000E+02
8.00000000E+01, 1.80000000E+02

8.05000000E+01, 0.00000000E+00
8.05000000E+01, 0.00000000E+00
8.05000000E+01, 0.00000000E+00
8.05000000E+01, 0.00000000E+00
7.95000000E+01, 1.80000000E+02
7.95000000E+01, 1.80000000E+02
7.95000000E+01, 1.80000000E+02
7.95000000E+01, 1.80000000E+02
7.95000000E+01, 1.80000000E+02
7.95000000E+01, 1.80000000E+02
8.05000000E+01, 0.00000000E+00
8.05000000E+01, 0.00000000E+00
8.05000000E+01, 0.00000000E+00
8.05000000E+01, 0.00000000E+00
7.95000000E+01, 1.80000000E+02
7.95000000E+01, 1.80000000E+02
7.95000000E+01, 1.80000000E+02
7.95000000E+01, 1.80000000E+02
7.95000000E+01, 1.80000000E+02
7.95000000E+01, 1.80000000E+02

8.10000000E+01, 0.00000000E+00
8.10000000E+01, 0.00000000E+00
8.10000000E+01, 0.00000000E+00
8.10000000E+01, 0.00000000E+00
7.90000000E+01, 1.80000000E+02
7.90000000E+01, 1.80000000E+02
7.90000000E+01, 1.80000000E+02
7.90000000E+01, 1.80000000E+02
7.90000000E+01, 1.80000000E+02
7.90000000E+01, 1.80000000E+02
8.10000000E+01, 0.00000000E+00
8.10000000E+01, 0.00000000E+00
8.10000000E+01, 0.00000000E+00
8.10000000E+01, 0.00000000E+00
7.90000000E+01, 1.80000000E+02
7.90000000E+01, 1.80000000E+02
7.90000000E+01, 1.80000000E+02
7.90000000E+01, 1.80000000E+02
7.90000000E+01, 1.80000000E+02
7.90000000E+01, 1.80000000E+02

8.15000000E+01, 0.00000000E+00
8.15000000E+01, 0.00000000E+00
8.15000000E+01, 0.00000000E+00
8.15000000E+01, 0.00000000E+00
7.85000000E+01, 1.80000000E+02
7.85000000E+01, 1.80000000E+02
7.85000000E+01, 1.80000000E+02
7.85000000E+01, 1.80000000E+02
7.85000000E+01, 1.80000000E+02
7.85000000E+01, 1.80000000E+02
8.15000000E+01, 0.00000000E+00
8.15000000E+01, 0.00000000E+00
8.15000000E+01, 0.00000000E+00
8.15000000E+01, 0.00000000E+00
7.85000000E+01, 1.80000000E+02
7.85000000E+01, 1.80000000E+02
7.85000000E+01, 1.80000000E+02
7.85000000E+01, 1.80000000E+02
7.85000000E+01, 1.80000000E+02
7.85000000E+01, 1.80000000E+02

8.20000000E+01, 0.00000000E+00
8.20000000E+01, 0.00000000E+00
8.20000000E+01, 0.00000000E+00
8.20000000E+01, 0.00000000E+00
7.80000000E+01, 1.80000000E+02
7.80000000E+01, 1.80000000E+02
7.80000000E+01, 1.80000000E+02
7.80000000E+01, 1.80000000E+02
7.80000000E+01, 1.80000000E+02
7.80000000E+01, 1.80000000E+02
8.20000000E+01, 0.00000000E+00
8.20000000E+01, 0.00000000E+00
8.20000000E+01, 0.00000000E+00
8.20000000E+01, 0.00000000E+00
7.80000000E+01, 1.80000000E+02
7.80000000E+01, 1.80000000E+02
7.80000000E+01, 1.80000000E+02
7.80000000E+01, 1.80000000E+02
7.80000000E+01, 1.80000000E+02
7.80000000E+01, 1.80000000E+02

8.25000000E+01, 0.00000000E+00
8.25000000E+01, 0.00000000E+00
8.25000000E+01, 0.00000000E+00
8.25000000E+01, 0.00000000E+00
7.75000000E+01, 1.80000000E+02
7.75000000E+01, 1.80000000E+02
7.75000000E+01, 1.80000000E+02
7.75000000E+01, 1.80000000E+02
7.75000000E+01, 1.80000000E+02
7.75000000E+01, 1.80000000E+02
8.25000000E+01, 0.00000000E+00
8.25000000E+01, 0.00000000E+00
8.25000000E+01, 0.00000000E+00
8.25000000E+01, 0.00000000E+00
7.75000000E+01, 1.80000000E+02
7.75000000E+01, 1.80000000E+02
7.75000000E+01, 1.80000000E+02
7.75000000E+01, 1.80000000E+02
7.75000000E+01, 1.80000000E+02
7.75000000E+01, 1.80000000E+02

8.30000000E+01, 0.00000000E+00
8.30000000E+01, 0.00000000E+00
8.30000000E+01, 0.00000000E+00
8.30000000E+01, 0.00000000E+00
7.70000000E+01, 1.80000000E+02
7.70000000E+01, 1.80000000E+02
7.70000000E+01, 1.80000000E+02
7.70000000E+01, 1.80000000E+02
7.70000000E+01, 1.80000000E+02
7.70000000E+01, 1.80000000E+02
8.30000000E+01, 0.00000000E+00
8.30000000E+01, 0.00000000E+00
8.30000000E+01, 0.00000000E+00
8.30000000E+01, 0.00000000E+00
7.70000000E+01, 1.80000000E+02
7.70000000E+01, 1.80000000E+02
7.70000000E+01, 1.80000000E+02
7.70000000E+01, 1.80000000E+02
7.70000000E+01, 1.80000000E+02
7.70000000E+01, 1.80000000E+02


8.35000000E+01, 0.00000000E+00
8.35000000E+01, 0.00000000E+00
8.35000000E+01, 0.00000000E+00
8.35000000E+01, 0.00000000E+00
7.65000000E+01, 1.80000000E+02
7.65000000E+01, 1.80000000E+02
7.65000000E+01, 1.80000000E+02
7.65000000E+01, 1.80000000E+02
7.65000000E+01, 1.80000000E+02
7.65000000E+01, 1.80000000E+02
8.35000000E+01, 0.00000000E+00
8.35000000E+01, 0.00000000E+00
8.35000000E+01, 0.00000000E+00
8.35000000E+01, 0.00000000E+00
7.65000000E+01, 1.80000000E+02
7.65000000E+01, 1.80000000E+02
7.65000000E+01, 1.80000000E+02
7.65000000E+01, 1.80000000E+02
7.65000000E+01, 1.80000000E+02
7.65000000E+01, 1.80000000E+02

##END
\end{lstlisting}

\end{document}